# Superconductivity and phase instability of NH$_3$-free Na-intercalated FeSe$_{1-z}$S$_z$

Jiangang Guo,[1,†] Hechang Lei,[1,†] Fumitaka Hayashi,[1] Hideo Hosono[1,2,3,*]

[1] Frontier Research Center, Tokyo Institute of Technology, Yokohama 226-8503, Japan

[2] Materials Research Center for Element Strategy, Tokyo Institute of Technology, Yokohama 226-8503, Japan

[3] Materials and Structures Laboratory, Tokyo Institute of Technology, Yokohama 226-8503, Japan

**The discovery of ThCr$_2$Si$_2$-type A$_x$Fe$_{2-y}$Se$_2$ (A = K, Rb, Cs and Tl) with $T_c$ ~ 30K make much progress in iron-based superconducting field, but their multiple-phase separations are disadvantageous for understanding the origin. On the other hand, for small alkali metals, studies on (Li,Na)FeCu(S,Se)$_2$ and NaFe$_{2-\delta}$S$_2$ shows that these compounds possess CaAl$_2$Si$_2$-type structure, implying that ThCr$_2$Si$_2$-type structure is unstable for small alkali-metal intercalated FeSe under high-temperature. Here we report a new intercalate Na$_{0.65(1)}$Fe$_{1.93(1)}$Se$_2$ with $T_c$ ~ 37 K, synthesized by low-temperature ammonothermal method. The notable finding is that the Na$_{0.65(1)}$Fe$_{1.93(1)}$Se$_2$ shows a ThCr$_2$Si$_2$-type structure, which is the first instance of small-sized alkali metal intercalates without NH$_3$ co-intercalation. Besides, the NH$_3$-poor Na$_{0.80(4)}$(NH$_3$)$_{0.60}$Fe$_{1.86(1)}$Se$_2$ and NH$_3$-rich phase with $T_c$s at 45 K and 42 K are identified by tuning the concentration of Na-NH$_3$ solutions. The modulation of interlayer spacing reveals the versatile evolution of structural stability and superconductivity in these intercalates.**

# Introduction

More recently, the structurally-simplest FeSe with space group (S. G.) *P4/nmm* has become very attractive since its $T_c$ (8 K) can be drastically enhanced by a factor of 5 under external pressure [1,2]. Moreover, the single-layer FeSe film with huge superconducting gap (~ 20 meV) implies that the potential $T_c$ could reach as high as 65 K even its bulk superconductivity (SC) decreases to ~ 40 K[3]. Another high-$T_c$ bulk SC derived from FeSe is intercalated $A_xFe_{2-y}Se_2$, which is synthesized at high temperature (~ 1300 K) but is only available for the large-sized alkali metals (A = K, Rb, Cs and Tl) [4,5,6,7]. The bulk $T_c$ of $A_xFe_{2-y}Se_2$ is ~ 30 K and the average crystal structure is a body-center tetragonal phase (S. G. *I4/mmm*)[4]. However, the origin of SC and the precise superconducting composition are still under debate due to the intrinsic phase separation and inhomogeneity in these materials[8,9,10].

The low-temperature solution method is another way to approach the intercalated phase, which is broadly applied and suitable for intercalating alkaline and alkali-earth metals even those with small ionic radius. Many superconductors such as $A_xC_{60}$, and $A_xMNX$ (A = Li-K, Ca-Ba, Yb and Eu; M = Ti, Zr, and Hf; X = Cl, Br, and I) were obtained through this method[11,12]. The relatively mild reaction keeps the host structure intact; therefore, the pure charge-transfer without destroying the conductive layer might favor the higher $T_c$. Actually, the FeSe intercalates obtained from low-temperature alkali metal and $NH_3$ co-intercalation exhibited a higher $T_c$ of 30 ~ 46 K compared with the samples obtained from the high-temperature method[13]. One surprising feature is that small-sized cations, such as Li and Na, combined with the $[NH_2]^-$ anions or $NH_3$ molecules can be intercalated into the FeSe layers[13]. However, a closer examination of the discontinuity of separation between Fe layers (*d*), *i. e.* the optimal Li- (*d* ≈ 8.3 Å),

Na- ($d \approx 8.7$ Å) and K- ($d \approx 7.4$ Å) intercalates (see Table 1), implies that the small cations may have more diverse coordination environments and arrangements of ligand-groups compared with larger ones[14, 15]. Furthermore, the fairly unstable $NH_3$-rich phase $Li_{0.56(1)}N_{1.72(2)}D_{4.63(3)}Fe_2Se_2$ with bilayers of $NH_3$ molecules has been prepared at 250 K and its crystal structure is determined from the neutron powder diffraction pattern collected at 5 K[16]. However, this $NH_3$-rich phase has not been reported to intercalate with Na cations so far. These intricate complexations of alkali metals between the [$Fe_2Se_2$] interlayers imply that the cation-intercalation potentially induces versatile intercalate especially for small alkali-metals, and need to be carefully investigated.

In this communication, we identify three Na/$NH_3$-intercalated phases from host FeSe through tuning the concentration of the Na-$NH_3$ solution and subsequent post-evacuation. The $T_c$s of 37 K, 45 K and 42 K correspond to $NH_3$-free $Na_{0.65(1)}Fe_{1.93(1)}Se_2$ (Phase I), $NH_3$-poor $Na_{0.80(4)}(NH_3)_{0.60(1)}Fe_{1.86(1)}Se_2$ (Phase II) and $NH_3$-rich phase (Phase III), respectively. The $NH_3$-free phase with $ThCr_2Si_2$-type structure is the first reported small alkali-metal intercalated iron selenide without accompanying $NH_3$ molecules. Moreover, the substitution of S for Se demonstrates that the structural stability and SC of this $ThCr_2Si_2$-type phase are more sensitive to the modulation of $d$ than the $NH_3$-poor intercalate.

**Results**

**Multiple phases in Na-intercalated FeSe superconductors.** Figure 1(a) shows the PXRD patterns for Na-intercalated FeSe phases prepared by immersion in 0.03-, 0.1-, and 0.3-mol L$^{-1}$ Na-$NH_3$ solutions. The phase I was obtained by immersion of FeSe in 0.3-mol L$^{-1}$ Na-$NH_3$ solution and subsequent post-evacuation treatment at ~ 260 K. As shown in Fig. 1(a), there are three distinct phases judged by the differently strongest peak positions of (00*l*) diffraction, which locate at 12.87°, 10.06° and 7.98° for the three phases, respectively. Fitting these patterns with highly preferred orientation, the separations between Fe layers (*d*) are 6.8339(2) Å, 8.7082(2) Å and 11.0721(3) Å, respectively. The schematic structure of the three intercalates with different $NH_3$ contents and *d* values are shown in Fig. 1(b). It can been that their *d* values increase from phase I to phase III patterns, indicating that the amount of intercalated cations or molecular groups between the FeSe layers also increases.

We analyzed the chemical compositions of phase I-III compounds using an electron-probe micro-analysis (EPMA) and ion chromatography (IC) methods. The EPMA analysis shows that the composition of typical grains for phase I is $Na_{0.65(1)}Fe_{1.93(1)}Se_2$. The difference in *d* values between phase I and phase II is 1.8625 Å, which is consistent with that of $AE_xHfNCl$ and $AE_x(NH_3)_yHfNCl$ (AE = Ca, Sr) (Ref. 17). Since the detected content of nitrogen is very small (~0.1), $Na_{0.65(1)}Fe_{1.93(1)}Se_2$ is called the $NH_3$-free phase. Phase II is obtained from 0.1-mol $L^{-1}$ $Na-NH_3$ solutions, which is consistent with the reported intercalates[13] and is called the $NH_3$-poor phase. The results of EPMA and IC analyses indicated that the composition of phase II was $Na_{0.80(4)}(NH_3)_{0.60}Fe_{1.86(1)}Se_2$, assuming that all nitrogen species came from $NH_3$ molecules (Supplementary Figure 1). The quite small difference in the Fe content between phase I and II may come from the reasonable uncertainties such as the highly chemical instability. It is noted that the formal iron valences of $Na_{0.65(1)}Fe_{1.93(1)}Se_2$, +1.73(1), is significantly lower than normal Fe valence (+2) in FeSe. Here we also try to analyze the nitrogen species using Raman and FT-IR techniques, but we could not identify whether the nitrogen species was an ammonia molecule or amide group (not shown). Therefore, the Fe oxidation state of $NH_3$-contaning intercalates cannot be determined. Note that the phase II easily decomposes even under inert conditions ($O_2$, $H_2O$ conc., < 0.1 ppm) at room temperature, which is attributed to the deintercalation of $NH_3$ molecules from the interlayers. Furthermore, the difference in $d$ between phase II and phase III is 2.369 Å, suggesting that phase III contains more $NH_3$ molecules. The larger $d$ of phase III could be attributed to the hydrogen bonding and additional intermolecular repulsion of $NH_3$. Thus, it is called the $NH_3$-rich phase. It should be noted that this $NH_3$-rich phase rapidly decomposes even at 250 K. It suggests that the $NH_3$-rich phase is quite hard to purify by conventional methods similarly as the Li-intercalated counterpart[16]. Possible reaction schemes for the three intercalated phases are discussed in detail (Supplementary Discussion).

Figure 1(c) shows the superconducting transitions of three samples in the lower temperature range. Distinct superconducting transitions are observed, where the $T_c$s of 37 K and 45 K correspond to the $NH_3$-free and $NH_3$-poor intercalates, respectively. The high superconducting volume fractions, 80 ~ 100%, and the M-H loops of 2 K (Supplementary Figure 3) suggest that both phases are bulk SC. The magnetization curve of the $NH_3$-rich phase shows two transitions of 45 K and 42 K, which are consistent with the features of mixture phases. Thus, the latter $T_c$ should be assigned to the new $NH_3$-rich phase. This fact reveals that the $T_c$ is not simply proportionate to the separation between neighboring FeSe layers[18]; the amount of charge-transfer, $FeSe_4$ tetrahedron distortion and interlayer coupling should also seriously influence the SC.

**Crystal Structure.** The experimental chemical composition of $NH_3$-free $Na_{0.65(1)}Fe_{1.93(1)}Se_2$ indicates that each layer contains 3.5% Fe-vacancies. The number of Fe-vacancies is strikingly smaller than that of high-temperature-synthesized $K_xFe_{2-y}Se_2$ phase. On the other hand, the abnormal valence of Fe implies that $Na_{0.65(1)}Fe_{1.93(1)}Se_2$ may be a metastable phase. To exactly determine the crystal structure of the $NH_3$-free and $NH_3$-poor compounds, we measured the PXRD patterns with a capillary assembled on a Mo-K$_a$ anode diffractometer. The PXRD pattern was collected at room temperature and the Rietveld refinement result is plotted in Fig. 2(a). The crystal structure of $Na_{0.65(1)}Fe_{1.93(1)}Se_2$ is refined as an analogous $K_xFe_{2-y}Se_2$ structural model[4]. The refinements smoothly converge to $R_p$ = 3.82 %, $R_{wp}$ = 4.32% and GOF = 1.90, respectively (Supplementary Table 1). The obtained crystal structure of $Na_{0.65(1)}Fe_{1.93(1)}Se_2$ has a body-centered tetragonal lattice (S. G., $I4/mmm$) with $a$ = 3.7870(4) Å, $c$ = 13.6678(4) Å and $V$ = 196.03(8) Å$^3$, reasonably smaller than those of $A_xFe_{2-y}Se_2$ (A=K, Rb, Cs). The Fe-Fe distance of 2.6778(3) Å is comparable with those of other $A_xFe_{2-y}Se_2$ phases. The Fe-Se distance is 2.4692(3) Å, which is the largest value among the pure FeSe as well as the intercalated phases. Meanwhile, the Se-Fe-Se angle of 100.13(2)° (×2) and 114.33(4)° (×4) indicates its FeSe$_4$ tetragonal bears rather large distortion compared with other intercalated phases.

Furthermore, since the value of $d$ in present $NH_3$-poor phase is comparable to that of $Li_{0.6}N_{1.0}D_{2.8}Fe_2Se_2$ (Ref. 14), we tried to use the reported crystal structure of $Li_{0.6}N_{1.0}D_{2.8}Fe_2Se_2$ to refine the PXRD pattern. The Rietveld refinement of the PXRD pattern of $Na_{0.80(4)}(NH_3)_{0.60}Fe_{1.86(1)}Se_2$ intercalates is shown in Fig. 2(b). Although the position of H cannot be obtained due to the smaller scattering factor of hydrogen, we could determine the unique variable of the Se position (4 $e$ site: (0, 0, z)) in this high-symmetry unit cell (Supplementary Table 2). The refined lattice parameters are $a$ = 3.7991(2) Å, $c$ = 17.4165(4) Å and $V$ = 251.38(6) Å$^3$, reasonably larger than those of $Li_{0.6}N_{1.0}D_{2.8}Fe_2Se_2$. The Fe-Se bond length is 2.4110(2) Å, which is comparable to that of FeSe counterparts, indicating that presence of $NH_3$ molecules weakens the structural change of FeSe layer.

**Structure instability in Na-intercalated Fe(Se,S).** Our experiments show that the intercalation of the smallest alkali metal, Li, does not yield the $NH_3$-free $ThCr_2Si_2$-type phase. This result suggests that the large $d$ of [$Fe_2Se_2$] layers could not hold the small Li without combining [$NH_2$]$^-$ anions or $NH_3$ molecules. As the previously reported[19,20,21,22], the smaller alkali metals intercalated into [$FeCuSe_2$] layers are more likely to form the $CaAl_2Si_2$-type structure with hexagonal-close-packed structure instead of $ThCr_2Si_2$-type. Besides, it was found the $NH_3$-free $ThCr_2Si_2$-type phase cannot be formed when the solid solution $FeSe_{1-z}Te_z$ act as the host compounds. This fact indicates that the enlarged $d$ in the host also cannot hold the small Na cation and destabilizes the $ThCr_2Si_2$ phase. It is likely that the $Na_{0.65(1)}Fe_{1.93(1)}Se_2$ phase resembles the metastable iron pnictide superconductor $Na_{1-y}Fe_{2-x}As_2$, which is on the edge of structural stability[23]. It is also known that the isovalent substitution of S for Se site could effectively tune the scale of $d$ and the SC of FeSe (Ref. 24) and $K_xSe_{2-y}Se_2$ (Ref. 25). Therefore, S-substituted $Na_{0.65(1)}Fe_{1.93(1)}(Se_{1-z}S_z)_2$ and $NH_3$-poor $Na_{0.80(4)}(NH_3)_{0.60}Fe_{1.86(1)}(Se_{1-z}S_z)_2$ samples were synthesized to investigate their structural and SC properties responses to the reduced $d$. Since the tetragonal phase of $FeSe_{1-z}S_z$ vanishes as $z$ rises above 0.5, the present work only focus on the low-S range ($0 \leq z \leq 0.5$). The $T_c$ of $FeSe_{1-z}S_z$ first increases to 12 K and then rapidly decreases to zero (Supplementary Figure 2), what is consistent with the previous report[24].

The PXRD patterns of all S-substituted intercalated phases were measured (Supplementary Figure 3). It can be seen that the main phase is an intercalated phase, and the peak position of (002) diffraction shifts to a higher angle as the sulfur content increases, indicating the contraction of the unit cell. The refined lattice parameters of both S-substituted samples are plotted in Fig. 3(a). The linear decreases of *a*- and *c*-axes are clearly observed as z increases up to 0.5, which unambiguously demonstrates that the S atoms are substituted into the Se sites. It can be seen that the solubility limit for $Na_{0.65(1)}Fe_{1.93(1)}(Se_{1-z}S_z)_2$ phase is $z = 0.3$, which is significantly smaller than that of the $NH_3$-poor intercalates. Even when the host is tetragonal phase as $z =0.4$ and 0.5, we still cannot obtain $ThCr_2Si_2$-type phase, which only survives as the separation of Fe layer is larger than 6.6364(3) Å in $Na_{0.65(1)}Fe_{1.93(1)}(Se_{0.7}S_{0.3})_2$.

**Superconductivity evolution in Na-intercalated Fe(Se,S).** Figure 3(b) plots the magnetization curves of S-substituted samples at a magnetic field of 10 Oe. As z increases up to a maximum, the $T_c$ and shielding volume fractions of both intercalated phases are gradually suppressed by S substitution. For example, the $T_c$s are 25K for both end-members $Na_{0.65(1)}Fe_{1.93(1)}(Se_{0.7}S_{0.3})_2$ and $Na_{0.80(4)}(NH_3)_{0.6}Fe_{1.86(1)}(Se_{0.5}S_{0.5})_2$. The evolution of the SC and phase boundary of two intercalates and host FeSe as the function of S content are summarized in Fig. 4. It can be seen that the decrease of $T_c$ for the two intercalates is rather mild and similar to the trend of $T_c$ in $K_xSe_{2-y}(Se_{1-z}S_z)_2$ phase, where $T_c$ vanishes as the content of S rises to 0.8[25]. The percentage reductions of $T_c$ for $Na_{0.80(4)}(NH_3)_{0.60}Fe_{1.86(1)}(Se_{0.7}S_{0.3})_2$ and $Na_{0.65(1)}Fe_{1.93(1)}(Se_{0.7}S_{0.3})_2$ are 13% and 30%, indicating that the effect of S-substitution for the $NH_3$-poor phase is weaker than that for the $NH_3$-free phase. In the present work, the formation of higher-sulfur intercalates is seriously hampered by the emergence of hexagonal phase in the host compounds. The fabrication of higher-sulfur tetragonal $FeSe_{1-z}S_z$ with a soft chemical method is underway so as to fully understand the phase diagram.

**Discussion**

It is reported that the $ThCr_2Si_2$-type structure strongly competes with the CaAl$_2$Si$_2$-type, where the small alkali-metal intercalation and the smaller $d$ would destabilize the ThCr$_2$Si$_2$-type[26]. Moreover, if the Se was completely substituted by S through high-temperature treatment, the product would be the CaAl$_2$Si$_2$-type NaFe$_{1.6}$S$_2$. The phonon spectrum calculation of hypothetical ThCr$_2$Si$_2$-type NaFe$_{1.6}$S$_2$ indicates that negative vibration frequencies of Na atoms result in structural instability[22]. In present case, it is found that the Na-Na distance (~ 3.79 Å) in Na$_{0.65(1)}$Fe$_{1.93(1)}$Se$_2$ is rather short than the counterpart (~ 3.86 Å) of CaAl$_2$Si$_2$-type NaFe$_{1.6}$S$_2$. With increasing the content of S, both Na-Na distances of intercalates further decrease to ~3.76 Å for $z$ = 0.3. The shortened Na-Na distance would inevitably increase Na-Na Coulomb repulsion, which may be the origin of structural destabilization. As $z$ = 0.3, the shortest Na-Na distance in Na$_{0.65(1)}$Fe$_{1.93(1)}$Se$_2$ induces the strongest Na-Na repulsion, which possibly induces the collapse of ThCr$_2$Si$_2$-structure. Therefore, our results show that the formation of the ThCr$_2$Si$_2$-phase for NH$_3$-free Na-intercalated FeSe requires a stricter chemical environment and the unique treatment procedure compared with the NH$_3$-containing intercalates.

Another point should be noted is the different $T_c$ and their evolutions between host and intercalates. It is fact that the $T_c$, to some extent, will increase as $d$ value increase from 5.52 Å to 8.71 Å. But $T_c$ decreases slightly again as $d$ above 8.71 Å. Actually, the summarized data on FeSe-based superconductors shows there is indeed an optimal $d$ value, ~ 8.6 Å, below which the interlayer spacing is in proportion to $T_c$[27]. As $d$ beyond this optimal value or fluctuates in a narrower range, other factors such as FeSe$_4$ distortion, Se height off Fe layers and transferred charge from Na to FeSe layer, would seriously influence the value and evolution of $T_c$.

On the other hand, the suppression of SC by chemical pressure is quite similar to that the pressured LiFeAs, where the shrinkage of unit cells leads to the monotonous decrease of $T_c$ and shielding volume fractions[28]. Therefore, it is highly expected that the full range S/Te-substituted intercalates should be prepared so as to thoroughly study the evolution of SC. Moreover, this ThCr$_2$Si$_2$-type Na$_{0.65(1)}$Fe$_{1.93(1)}$Se$_2$ contains very few Fe-vacancies and significantly differs from the highly Fe-vacant A$_x$Fe$_{2-y}$Se$_2$ obtained at high temperatures. We expect that the discoveries of intercalates with a nearly intact FeSe layer could promote the further understanding of ThCr$_2$Si$_2$-phase SC.

## Methods

**Synthesis.** The high-purity powder precursors were synthesized using a modified high-temperature solid-state method. The typical process of FeSe and its solid solution uses iron granules (Alfa, 99.98%), selenium grains (Kojundo, 99.99%) and sulfur grains (Kojundo, 99.99%), which are put into alumina crucibles and sealed in silica ampoules. The samples were heated to 1300 K for 30 h, then annealed at 673 K for 50 h, and finally furnace-cooled to room temperature. The Na pieces and Fe(Se, S) powders with nominal 1 : 2 mole ratio were loaded and sealed into a Taiatsu Glass TVS-N$_2$ high-pressure vessel with a magnetic stirrer. These manipulations were carried out in an argon-filled glove box with an O$_2$ and H$_2$O content below 1 ppm. The vessel was taken out from the glove box and connected to a vacuum/NH$_3$ gas line equipped with a turbo molecular pump and mass-flow controller. The vessel was evacuated using molecular pump firstly and then placed in a bath of ethanol cooled by liquid nitrogen (~ 223 K). The ammonia cylinder and regulator were then opened allowing ammonia to condense into the vessel. Typically, 4-6 g of NH$_3$ was condensed to form three different Na/NH$_3$ concentration solutions. Then, the reaction vessel was closed and stirred for 3 h at 223 K ~ 243 K. After the intercalated procedure finished, the vessel was opened and the solutions were evaporated at an ambient pressure. For NH$_3$-free sample, the vessel was further evacuated to ~ 10$^{-3}$ Pa using a molecular pump, whereas for other two NH$_3$-containing samples, the evacuation process was not needed.

**Structural and Magnetic characterization.** The powder x-ray diffraction (PXRD) patterns of products were measured by Bruker diffractometer model D8 ADVANCE with Cu-K$_\alpha$ anode ($\lambda$ = 1.5408 Å). To reduce the preferred orientation of PXRD patterns, the samples also were loaded into thin-walled capillary tubes and then measured using a Bruker diffractometer with Mo-K$_\alpha$ radiations ($\lambda$ = 0.7107 Å). The Rietveld refinement of patterns was performed using code TOPAS4 (TOPAS 2005, Version 3, Bruker AXS, Karlsruhe, Germany). The dc magnetization was measured by a vibrating sample magnetometer (SVSM, Quantum Design) at the low magnetic field of 10 Oe.

**Chemical composition characterization.** The chemical compositions of the samples were determined by electron probe microscope analysis (EPMA) with a backscattered electron (BSE) mode. The real composition was determined as the average value of 10 points on a typical grain with dimensions of 50 μm × 30 μm × 5 μm. The Na and nitrogen contents in the samples were determined using the ion chromatography (IC) technique[29]. Typically, about 10 mg of the sample was dissolved in 5 mol L$^{-1}$ HF aqueous solutions, and was diluted by adding water. The resultant solution containing Na$^+$ and NH$_4^+$ was analyzed by IC with a Shimadzu CDD-10A conductivity detector.

# References

1. Hsu, F. C. *et al*. Superconductivity in the PbO-type structure *α*-FeSe. *Proc. Natl Acad.*

*Sci.* **105**, 14262-14264 (2008).

2. Medvedev, S. *et al*. Electronic and magnetic phase diagram of *β*-Fe$_{1.01}$Se with superconductivity at 36.7 K under pressure. *Nat. Mater.* **8**, 630-633 (2010).

3. Wang, Q. Y. *et al*. Interface-induced high-temperature superconductivity in single unit-cell FeSe films on SrTiO$_3$. *Chin. Phys. Lett.* **29**, 037402 (2012).

4. Guo, J. G. *et al*. Superconductivity in the iron selenide K$_x$Fe$_2$Se$_2$ ($0 \leq x \leq 1.0$). *Phys. Rev. B* **82**, 180520(R) (2010).

5. Krzton-Maziopa, A. *et al*. Synthesis and crystal growth of Cs$_{0.8}$(FeSe$_{0.98}$)$_2$: a new iron-based superconductor with $T_c$ = 27 K. *J. Phys.: Condens. Matter* **23**, 052203 (2011).

6. Wang, A. F. *et al.* Superconductivity at 32 K in single-crystalline Rb$_x$Fe$_{2-y}$Se$_2$. *Phys. Rev. B* **83**, 060512(R) (2010).

7. Fang, M. H. *et al.* Fe-based superconductivity with $T_c$ = 31 K bordering an antiferromagnetic insulator in (Tl, K) Fe$_x$Se$_2$. *Europhy. Lett.* **94**, 27009 (2011).

8. Ding, X. X. *et al*. Influence of microstructure on superconductivity in K$_x$Fe$_{2-y}$Se$_2$ and evidence for a new parent phase K$_2$Fe$_7$Se$_8$. *Nat. Comm.* **4**, 1897 (2013).

9. Zhao, J., Cao, H. B., Bourret-Courchesne, E. D., Lee, D. H. & Birgeneau, R. J. Neutron-diffraction measurements of an antiferromagnetic semiconducting phase in the vicinity of the high-temperature superconducting state of K$_x$Fe$_{2-y}$Se$_2$. *Phys. Rev. Lett.* **109**, 267003 (2012).

10. Zhang, A. M. *et al*. Superconductivity at 44 K in K intercalated FeSe system with excess Fe. *Sci. Rep.* **3**, 1216 (2013).

11. Buffinger, D. R., Ziebarth, R. P., Stenger, V. A., Recchia, C. & Pennington, C. H.

Rapid and efficient synthesis of alkalimetal-C$_{60}$ compounds in liquid ammonia. *J. Am. Chem. Soc.* **115**, 9267–9270 (1993).

12. Yamanaka, S. *et al*. Preparation and superconductivity of intercalation compounds of TiNCl with aliphatic amines. *J. Mater. Chem.* **22**, 10752-10762 (2012).

13. Ying, T. P. *et al*. Observation of superconductivity at 30 ~ 46 K in A$_x$Fe$_2$Se$_2$ (A = Li, Na, Ba, Sr, Ca, Yb, and Eu). *Sci. Rep.* **2**, 426 (2012).

14. Burrard-Lucas, M. *et al*. Enhancement of the superconducting transition temperature of FeSe by intercalation of a molecular spacer layer. *Nat. Mater.* **12**, 15 (2013).

15. Ying, T. P. *et al*. Superconducting phases in potassium-intercalated iron selenides. *J. Am. Chem. Soc.* **135**, 2951-2954 (2013).

16. Sedlmaier, S. J. *et al*. Ammonia-rich high-temperature superconducting intercalates of iron selenide revealed through time-resolved in situ x-ray and neutron diffraction. *J. Am. Chem. Soc.* **136**, 630-633 (2014).

17. Zhang, S. *et al*. Superconductivity of alkali metal intercalated TiNBr with α-type nitride layers. *Supercond. Sci. Technol.* **26**, 122001 (2013).

18. Zheng, L. *et al.* Superconductivity in (NH$_3$)$_y$Cs$_{0.4}$FeSe. *Phys. Rev. B* **88**, 094521 (2013).

19. Llanos, J., Contreras-Ortega, C. & Mujica, C. Structure refinement of a new sulfide of copper and iron with layered structure. *Mater. Res. Bull.* **28**, 39-44 (1993).

20. Llanos, J., Contreras-Ortega, C., Paez, M., Guzman, M., & Mujica, C. Synthesis and structural characterization of two intercalated lithium and sodium copper iron selenides: LiCuFeSe$_2$ and NaCuFeSe$_2$. *J. Alloys Comp.* **201**, 103-104 (1993).

21. Fong, R., Dahn, J. R., Batchelor, R. J., Einstein, F. W. B. & Jones, C. H. W. New

$Li_{2-x}Cu_xFeS_2$ ($0 \leq x \leq 1$) and $Cu_xFeS_2$ (~$0.25 \leq x \leq 1$) phases. *Phys. Rev. B* **39**, 4424-4429 (1989).

22. Lai, X. *et al*. New layered iron sulfide $NaFe_{1.6}S_2$: synthesis and characterization. *Inorg. Chem.* **52**, 12860-12862 (2013).

23. Friederichs, G. M. *et al*. Metastable 11 K Superconductor $Na_{1-y}Fe_{2-x}As_2$. *Inorg. Chem.* **51**, 8161-8167 (2012).

24. Mizuguchi, Y., Tomioak, F., Tsuda, S., Yamaguchi, T. & Takano, Y. Substitution effects on FeSe Superconductor, *J. Phys. Soc. Jpn.* **78**, 074712 (2009).

25. Lei, H. C. *et al*. Phase diagram of $K_xFe_{2-y}Se_{2-z}S_z$ and the suppression of its superconducting state by an $Fe_2$-Se/S tetrahedron distortion. *Phys. Rev. Lett.* **107**, 137002 (2011).

26. Khatun, M., Stoyko, S. S. & Mar, A. Quaternary arsenides $AM_{1.5}Tt_{0.5}As_2$ (A = Na, K, Rb; M = Zn, Cd; Tt = Si, Ge, Sn): size effects in $CaAl_2Si_2$-and $ThCr_2Si_2^-$ type structures. *Inorg. Chem.* **52**, 3148-3158 (2013).

27 . Noji, T. *et al*. Synthesis and post-annealing effects of alkaline-metal-ethylenediamine-intercalated superconductors $A_x(C_2H_8N_2)_yFe_{2-z}Se_2$ (A = Li, Na) with $T_c$ = 45 K, *Physica C* **In press**, DOI: 10.1016/j.physc.2014.01.007.

28. Pitcher, M. J. *et al*. Response of superconductivity and crystal Structure of LiFeAs to hydrostatic pressure. *J. Am. Chem. Soc.* **131**, 2986-2992 (2009).

29. Hayashi, F., Ishizu, K. & Iwamoto, M. Fast and almost complete nitridation of mesoporous silica MCM-41, with ammonia in a plug-flow reactor. *J. Am. Ceram. Soc.* **93**, 104-110 (2010).

## Acknowledgements

This work was supported by the Funding Program for World-Leading Innovative R&D on Science and Technology (FIRST) and MEXT Element Strategy Initiative to form a core research center, Japan.

**Author contributions**

H. H. provided strategy and advice for the material exploration. J. G. and H. L. performed the sample fabrication, measurements and fundamental data analysis. F. H. set up the apparatus for low-temperature ammonothermal experiments and analyzed the chemical composition of products using the IC technique. J. G. and H. H. wrote the manuscript based on discussion with all the authors.

**Additional information**

**Supplementary Information** accompanies this paper.

**Competing financial interests**: The authors declare that they have no competing financial interests.

† These authors contributed equally to this work.

*Author to whom correspondence should be addressed. E-mail: hosono@msl.titech.ac.jp

# Figure captions

**Figure 1. Three Na/NH$_3$-intercalated FeSe phases.** (a) The powder x-ray diffraction patterns (PXRD) of three intercalates with highly preferred orientation and host FeSe. The values of $l$ are specified as 1 (primitive lattice) or 2 (body center lattice). (b) The schematic view of FeSe and three Na-intercalated phases with different separations between nearest Fe layers ($d$). (c) The magnetization curves of three intercalates measured with the zero-field-cooling (ZFC) and field-cooling (FC) modes at $H = 10$ Oe.

**Figure 2. Powder x-ray diffraction (PXRD) patterns and their Rietveld refinements.** (a) Results of the NH$_3$-free Na$_{0.65(1)}$Fe$_{1.93(1)}$Se$_2$ phase. This pattern comes from three phases: Na$_{0.65(1)}$Fe$_{1.93(1)}$Se$_2$ (top, 91.3%), FeSe (middle, 3.5%), and Fe$_7$Se$_8$ (bottom, 5.2%). The inset shows the obtained crystal structure of ThCr$_2$Si$_2$-type Na$_{0.65(1)}$Fe$_{1.93(1)}$Se$_2$. (b) The PXRD pattern and Rietveld refinement results for the NH$_3$-poor Na$_{0.80(4)}$(NH$_3$)$_{0.6}$Fe$_{1.86(1)}$Se$_2$ phase. This pattern comes from three phases: Na$_{0.80(4)}$(NH$_3$)$_{0.6}$Fe$_{1.86(1)}$Se$_2$ (top, 95.4%), FeSe (middle, 2.4%), and Fe$_7$Se$_8$ (bottom, 2.2%). The inset shows the crystal structure of Na$_{0.80(4)}$N$_{0.6}$Fe$_{1.86(1)}$Se$_2$. The lower blue line shows the difference between experimental and calculated values.

**Figure 3. Lattice constants and magnetic susceptibility.** (a) The lattice constants of Na$_{0.65(1)}$Fe$_{1.93(1)}$Se$_2$ and Na$_{0.80(4)}$(NH$_3$)$_{0.6}$Fe$_{1.86(1)}$Se$_2$ vs. S-substitution. The error bar represents one standard deviation of uncertainty and is shown in the center of symbol. The solid lines are guides to the eye. (b) The magnetization curves in the lower temperature region for two series of S-substitution intercalates. For clarity, only the ZFC

curves are shown.

**Figure 4. The evolutions of superconductivity and phase boundary.** All of $T_c$ were determined from the magnetization curves. Two different phase boundaries are clearly observed.

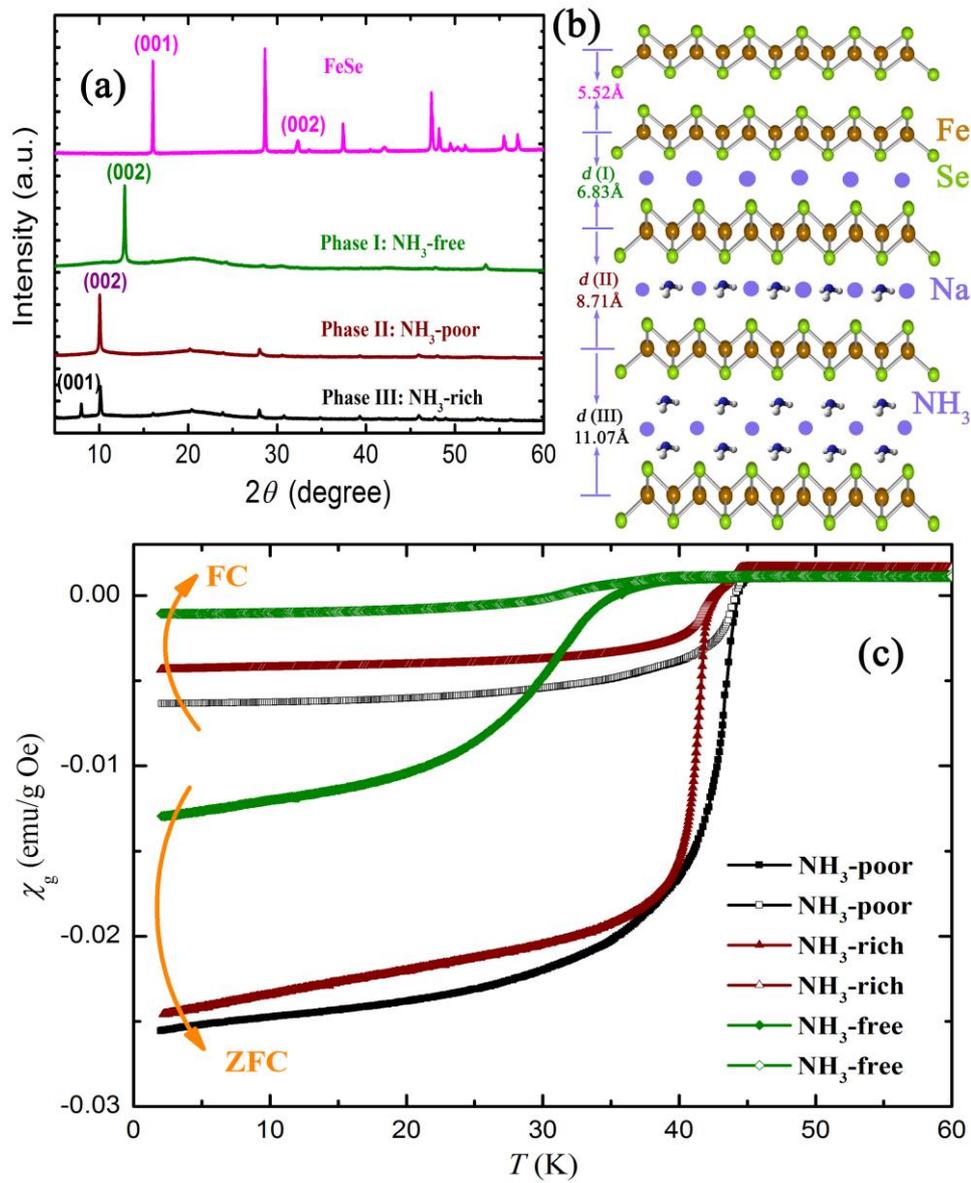

Fig. 1 *Guo* et al.

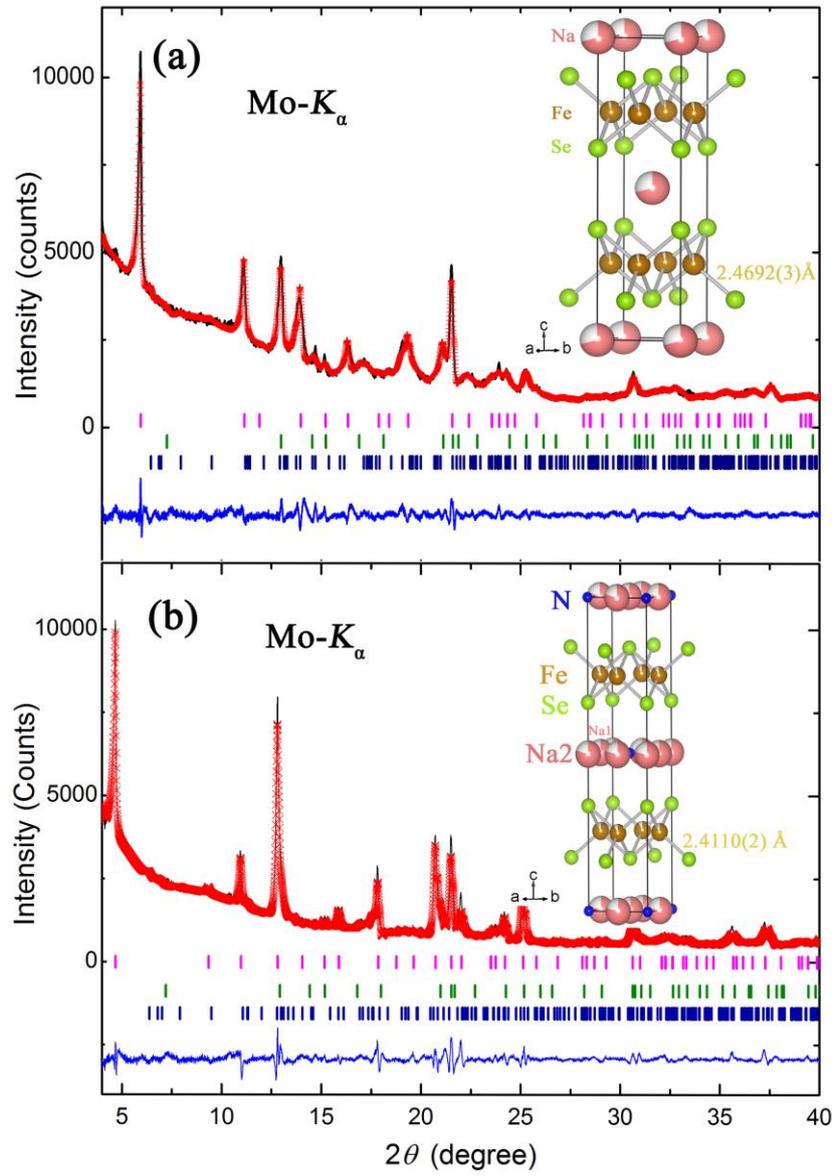

Fig. 2 *Guo* et al.

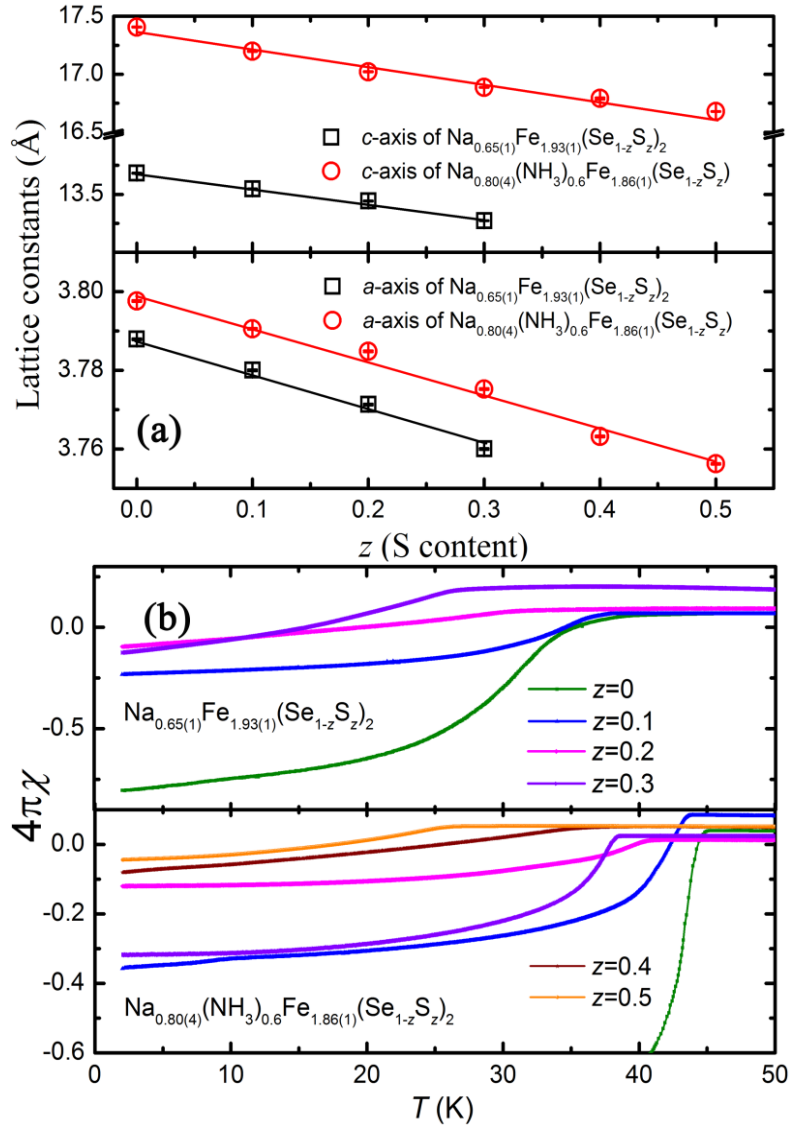

Fig. 3 *Guo* et al.

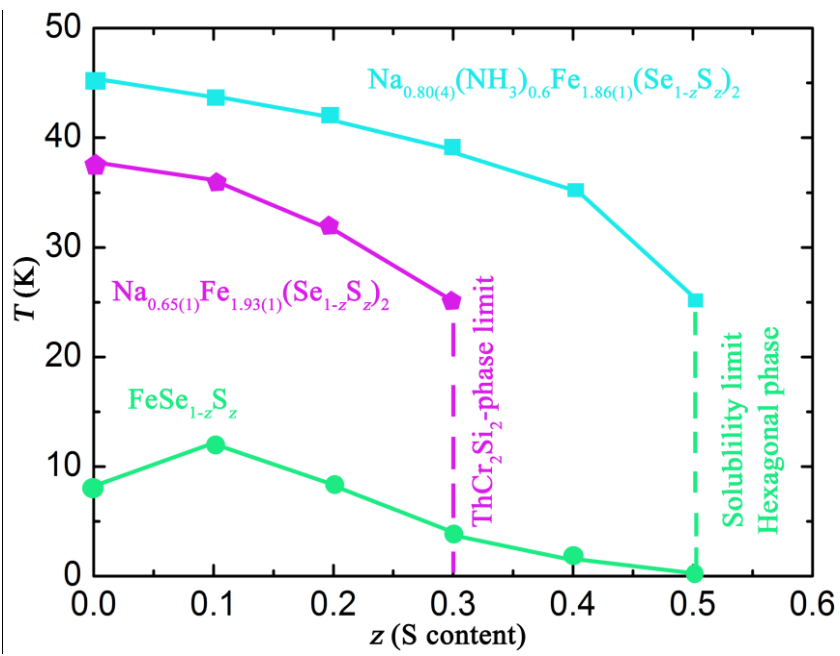

Fig. 4 *Guo* et al.

|  | NH$_3$-rich | | NH$_3$-poor | | | | NH$_3$-free | |
|---|---|---|---|---|---|---|---|---|
|  | | | Metal-poor | | Metal-rich | | | |
|  | $d$ | $T_c$ | $d$ | $T_c$ | $d$ | $T_c$ | $d$ | $T_c$ |
| Li$^+$ | ~ 9.0 | 39 | - | - | ~ 8.3 | 44 | - | - |
| Na$^+$ | ~ 11.1* | 42* | - | - | ~ 8.7 | 45 | ~ 6.8* | 37* |
| K$^+$ | ~ 10.2 | ? | ~ 7.8 | 44 | ~ 7.4 | 30 | - | - |
| AE$^{2+}$ | - | - | 8.0~8.4 | 35~40 | ~ 10.3 | 38~39 | - | - |
| RE$^{2+}$ | - | - | ~8.1 | 42 | ~ 10.2 | 40~42 | - | - |

*: This work  −: None  ?: Unknown

**Table 1: Intercalated FeSe superconducting phases.** The separation of nearest Fe layers $d$ (Å) and $T_c$ (K) of intercalated (A/AE/RE)$_x$(NH$_3$)$_y$(NH$_2$)$_z$Fe$_2$Se$_2$ (A = Alkali metals or AE = Alkali earth or RE = rare earth metals) superconductors synthesized by the ammonothermal method shown in literatures 13 - 16 and this work.